\documentclass[article]{svmono}

\usepackage[greek,english.german]{babel}
\usepackage{marvosym}
\usepackage{graphicx,caption}        %standard LaTeX graphics tool
\usepackage{booktabs}
\usepackage{threeparttable}

\def\begmarg{\par \begingroup  \leftskip2.5em \rightskip2em \large}
\def\endmarg{\par \endgroup }

\def\3H{{3\over2}}
\def\simlt{\lower.5ex\hbox{$\; \buildrel < \over \sim \;$}}
\def\simgt{\lower.5ex\hbox{$\; \buildrel > \over \sim \;$}}

\def\degs{$^{\circ}$}

\def\simlt{\lower.5ex\hbox{$\; \buildrel < \over \sim \;$}}
\def\simgt{\lower.5ex\hbox{$\; \buildrel > \over \sim \;$}}

\def\bcol{\begin{column}}
\def\ecol{\end{column}}
\def\bcols{\begin{columns}[]}
\def\ecols{\end{columns}}

\usepackage[german,short]{isodate}
\begin{document}

\pagestyle{plain}
\ \ 
\vspace{1cm}

\begin{center}
{\huge {\bf }}
\bigskip
  
{\LARGE {\bf JAN VETH'S PAINTINGS OF}} 
\bigskip

{\LARGE {\bf JACOBUS KAPTEYN}}
\vspace{1cm}

\noindent
{\Large Pieter C. van der Kruit},\\
{\large Kapteyn Astronomical Institute, University of Groningen,}\\
{\large P.O. Box 800, 9700AV Groningen, the Netherlands}.\\
{\large vdkruit@astro.Rug.nl; www.astro.rug.nl/$\sim$vdkruit}
\vspace{2cm}

\end{center}

{\large

\noindent
Version \today.
\vspace{2cm}

\noindent
%This manuscript has been submitted for publication
%to the {\it Journal of Astronomical History and
%Heritage}. This preprint version has been produced in \LaTeX.

\newpage

\noindent
{\Large {\bf Abstract}}\\
\bigskip

Jacobus C. Kapteyn was one of the most prominent astronomers worldwide in the beginning of the twentieth century and is nowadays regarded as one of the coryfees of the University of Groningen. His legacy is not only the prominence of Dutch astronomy during the twentieth century through his students like Jan Oort and Willem de Sitter and the Dutch school that followed. Part of his legacy is also the two paintings of him, produced in oil on canvas by prominent Dutch painter Jan Pieter Veth. One, showing him working at his desk, decorates the Kapteyn Room in the Kapteyn Astronomical Institute together with a painting of Mrs. Kapteyn by a different artist, and the other one, displaying Kapteyn in academic attire, is part of the University of Groningen’s gallery of professors in the Senate Chamber of the central Academy Building. The first was offered to Kapteyn and his wife on the occasion of his 40-th anniversary as professor in 1918 and the second to the University after his retirement in 1921.

There has been some confusion about the way in which these paintings have been produced, to the extent that it has been suggested that there must have been a third portrait that now is lost. Former director of the Kapteyn Astronomical Institute Adriaan Blaauw has proposed that the one in the Senate Chamber actually is a first version meant to be offered to Mrs. Kapteyn in 1918, but at her request replaced by the one now in the Kapteyn Room. The first version was then later adapted to the requirements of the gallery of professors by Veth himself by overpainting it with academic gown, jabot and beret. A preliminary trial version in oil on wood by Veth, in the possession of Kapteyn’s namesake and greatgrandson Jack Kapteyn, shows what this painting would have looked like before the adaption by Veth. 

Recently an exhibition of Veth’s work (including the two Kapteyn paintings) was held in the Dordrechts Museum, in Veth’s city of birth, where is was stated as  a fact that three paintings were produced of which one now is lost. The following reports on a critical evaluation of the available evidence, including the biography of Jan Veth that well-known historian Johan Huizinga, friend of Veth, wrote not long after the latter’s demise, and letters Veth wrote to his wife while he was working on these paintings in Groningen. I conclude that the evidence provides strong support of Blaauw’s proposed sequence of events with a few modifications, and that no third, now lost, painting has been produced.
\bigskip

\noindent
{\bf Key words:} History: Galaxy research, History: University of Groningen, History: Jacobus C. Kapteyn, Legacy: Painted portraits, Legacy: Jan P. Veth, Legacy: Professors Gallery

\newpage

\section{Introduction}\label{sect:Introduction}

One of the giants of Dutch astronomy, and in his days one of the most important astronomers in the world, is Jacobus Cornelius Kapteyn (1851–-1922). After his studies in mathematics, physics and astronomy in Utrecht and a brief spell at the Observatory of the University of Leiden, he was appointed professor at the University of Groningen in 1878. He remained in this position until his retirement in 1921, about a year before his death. Of the three state-financed universities, Leiden and Utrecht had well-equipped observatories that had been founded in the nineteenth century by respectively Frederik Kaiser (1808--1872), Professor of Astronomy in Leiden, and by  Christophorus Henricus Diedericus Buys Ballot (1817--1890), Professor of  Physics, particularly teaching meteorology and astronomy in Utrecht. Efforts of Kapteyn to obtain his own observatory failed, not because of lack of support from his own university but mostly due to opposition from the directors of the two other observatories, who were loath to share available resources with a third party. Therefore he founded his Astronomical Laboratory, where photographic material from other observatories was measured and interpreted. From this he rose to become one of the most influential astronomers in the world, being appointed in 1908 Research Associate of the Carnegie Institution of Washington and having his {\it Plan of Selected Areas} adopted by almost all major observatories to contribute, in particular as the primary observing program for the largest telescope in the world, the new 60-inch Telescope on Mount Wilson in California, when it became operational in that same year. For an extensive, academic biography of Kaptey see van der Kruit (2015), and for a more general one van der Kruit (2021a).

The University of Groningen rightly covets Kapteyn as one of its absolute champions. Not surprisingly he is depicted with only a few others in the large stained-glass windows that decorate the Aula, the Main Auditorium in the University’s Central Academy Building (see Fig.~14.10 in van der Kruit, 2015, or Fig.~10.7 in van der Kruit, 2021a, for a full view, of for a zoom in on Kapteyn's head Fig.~30 in van der Kruit, 2023). And more recently, in the year of the centenary of his death, 2022, Kapteyn was the second to be honored with a large wall painting in the center of Groningen, giving precedence only to the first female university student in the Netherlands. On the occasion, the Kapteyn Astronomical Institute published a bi-lingual booklet on the history of Groningen astronomy to commemorate Kapteyn (van der Kruit, 2022b).

The University of Groningen and the Kapteyn Astronomical Institute own two paintings of Kapteyn, both produced by one of the most celebrated portrait painters of the time, Jan Pieter  Veth (1862–1925). One has been presented by friends and colleagues at the celebration of his forty years' professorship in 1918 and a second one on the occasion of his retirement in 1921. The first of these depicts him working behind his desk and resides in the Kapteyn Room in the Kapteyn Astronomical Institute, where most of his books, his desk and some other belongings that decorated his office are kept. The second painting shows him dressed in academic gown, jabot (or bavette) and beret and is part of the extensive collection of paintings of professors in academic attire that decorates the Senate and more recently Faculty Chambers in the Academy  Building.

According to Adriaan Blaauw, Veth had originally produced a different painting to be presented in 1918, but this was rejected by the recipient Mrs. Kapteyn, because it did not depict him as she knew him, i.e. working behind his desk. The third director of the  Kapteyn Astronomical Laboratory, Adriaan Blaauw, has put forward the suggestion that the painting in the Senate Chamber is in fact the original 1918 one, overpainted by Veth himself with academic attire. The biography of Veth by Johan Huizinga and availability of online versions of Veth’s letters to his wife, when he was away from home working on the paintings, sheds new light on this. And these are the subject of this paper.

For those not familiar with the Dutch language I add that  the ‘h’ in Veth is silent and the name Veth is pronounced like ‘vet’, the  often used short for veteran or for veterinarian.

\section{Jacobus C. Kapteyn}\label{sect:JCK}

Jacobus Cornelius Kapteyn (Fig.~\ref{fig:JCK}) was Professor of Astronomy at the University of Groningen in the Netherlands from 1878 to 1921. In his days he was one of the most prominent astronomers in the world (van der Kruit \&\ van Berkel, 2000, van der Kruit, 2015, 2021a). His appointment in Groningen was a direct consequence of a thorough revision of higher education in the Netherlands in a law of 1876, in the spirit of Dutch liberal Prime Minister, Johan Rudolph Thorbecke (1798--1872), among others author of the 1848 revision of the Constitution, establishing the parliamentary democracy. This law stipulated, among others that curricula should be the same and therefore astronomy should be taught at all three national universities, opening up the possibility for an astronomy professorship in Groningen in addition to the ones in Leiden and Utrecht.

His life-long interest was the determination of the distribution of the stars in space. This required first and foremost good quality catalogues, which existed really only for the part of the sky visible from the northern hemisphere (the {\it Bonner Durchmusterung}). His first major effort was to supply a catalogue of equal quality for the southern hemisphere in a collaboration with David Gill (1843–1914), director of the Royal Observatory at the Cape of Good Hope. Gill had proposed to use photographic material for this and he provided the necessary photographic plates that were measured up in Groningen. Kapteyn worked for twelve years on this. The resulting {\it Cape Photographic Durchmusterung} contained almost half a million stars with positions and apparent magnitudes (brightnesses), published in three large volumes, the final one appearing in the last year of the nineteenth century (Gill \&\ Kapteyn, 1896, 1897, 1900).

\begin{figure}[t]
\sidecaption[t]
%\begin{center}
\includegraphics[width=0.64\textwidth]{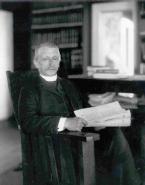}
%\end{center}
\caption{\normalsize   Jacobus Cornelius Kapteyn (1851--1922) was Professor of Astronomy, Probability Theory and Mechanics at the  University of Groningen between 1878 and his retirement in 1921, and from 1908 onwards Research Associate of the Carnegie Institution of Washington. From 1908 to 1914 he annually visited the Institution’s Mount Wilson Observatory near Pasadena (close to Los Angeles) to oversee progress in the Observatory’s contribution to his \textit{Plan of Selected Areas}. This photograph shows him in the library of the `Monastery’, the residence of observing astronomers, at Mount Wilson. Courtesy Kapteyn Astronomical Institute.}
\label{fig:JCK}
\end{figure}

But more was needed, particularly counts of fainter stars, but also measurements of the tiny displacement of stars on the sky on account of their motion in space (proper motions), that Kapteyn used to statistically estimate distances, at least for the brightest and the nearest stars. This could only be accomplished by a consorted effort of many observatories all around the world. Kapteyn designed an approach to this  that involved  observations of stars in each of 206 small areas across the whole sky. For the {\it Plan of Selected Areas} (Kapteyn, 1906) he ensured the cooperation of twenty major and some less prominent observatories around the  world  to send their observations, usually in the form of photographic plates, to Kapteyn in Groningen to be measured in his Astronomical Laboratory. George Ellery Hale (1868--1938) in 1908 adopted Kapteyn’s {\it Plan} as the primary program for his brand new, giant 60-inch telescope on Mount Wilson near Pasadena and Los Angeles, the largest in the world, and had Kapteyn appointed for life as a Research Associate by the Carnegie Institution of Washington. Starting in  1908  Kapteyn annually visited Mount Wilson up to 1914, when the First World War prevented him (and his wife, who accompanied him) from crossing the Atlantic. 

The {\it Plan} took a long time to complete and only a small part of the observations were carried out and reduced by the time of Kapteyn’s retirement in 1921. Using existing data, Kapteyn produced towards the end of his life a first model of our Galaxy (Kapteyn \&\ van Rhijn, 1920; Kapteyn, 1922).  Because the counts along the Milky Way on the sky showed little  variation, the Sun appeared to be near the center, while the decrease away from the Milky Way translated into a highly flattened structure. Kapteyn (1922) further pioneered the field of alactic dynamics, the interplay between structure (the distribution of stars and other matter) and kinematics (the motions). For the system to be stable the gravitational forces have to be in equilibrium with the motions, either in the form of a centrifugal force from rotation or from random  motions countering a tendency to collapse.  This `Kapteyn Universe’ was a complete and consistent model.

The dynamics in the vertical direction survived as valid and approximately correct up till the present. In the plane of the Milky Way there is much attenuation of the light due to scattering by intervening dust. Kapteyn had worried a long time about that — and even surmised it would be more  effective at blue wavelengths and therefore ‘redden’ the colors of distant stars —, but American astronomer Harlow Shapley (1885–1972) had incorrectly convinced him there was no such extinction. After his death definite evidence for large amounts of dust in the plane of the Milky Way was found and the stellar system was in reality much larger. A vital contribution to this development was presented by Kapteyn’s most famous student, Jan Hendrik Oort (1900–1992). Oort had studied with Kapteyn (van der Kruit, 2019, 2021b), but finished his PhD thesis after Kapteyn had died. He  found that the larger system rotated around a distant center (Oort, 1927). Oort also wrote the definitive paper of the {\it Plan}, showing signs of the phenomenon that the main part of the Galaxy, the disk, contained spiral structure as in many external galaxies (Oort, 1938). Kapteyn and later Oort were among the very select few most prominent astronomers worldwide of their times, together with Marcel Gilles Jozef Minnaert (1893--1970) in Utrecht and Antonie Pannekoek (1873--1960) in Amsterdam laying the foundations for the current prominence of Dutch astronomy.

That Kapteyn is regarded as one of the most prominent professors of the university of Groningen, is also evident from the fact that on three occasions, including the tricentennial in 1914, he was allowed to propose honorary doctorates to be awarded by the university (van der Kruit, 2021c). His importance was also the basis of a {\it Legacy Symposium} about him in 1999 on the occasion of the University of Groningen's 385-th anniversary (van der Kruit \&\ van Berkel, 2000). Following up on this I published  a comprehensive, academic  biography of Jacobus Kapteyn (van der Kruit, 2015) and a more general audience version (van der Kruit, 2021a). Further information on the prominence of Dutch Galactic astronomy, building on Kapteyn, can be found in my quite similar set of biographies of Jan Oort (van der Kruit, 2019, 2021b) and a paper in this journal on Kapteyn’s successor Pieter van Rhijn (van der Kruit, 2022a).

\section{Jan P. Veth}\label{section:Veth}

For this section I have drawn extensively on three authoritative publications on Veth. The first is the biography by famous historian Johan Huizinga (1927), who had known him well. The book by Fusien Bijl de Vroe (1987) is a very detailed description of Veth’s life based mainly on citations from his letters and illustrated with many of his drawings and paintings.  She is an art historian, but in addition a greatgranddaughter of Veth. Bijl de Vroe {\it et al.} (2023) is a catalogue accompanying a major exhibition of Veth’s paintings in the Dordrechts Museum in 2023. Dordrecht, some 20 km southeast of Rotterdam,  is the city of his birth. The exhibition displayed the two paintings of Jacobus C. Kapteyn that are part of the collection of the University of Groningen and the Kapteyn Astronomical Institute. 
\bigskip

\begin{figure}[t]
\sidecaption[t]
%\begin{center}
\includegraphics[width=0.64\textwidth]{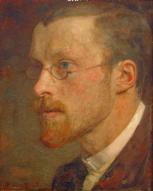}
%\end{center}
\caption{\normalsize Self-portrait of Jan Veth, painted in 1887. Oil on panel, $35 \times 26$ cm, Dordrechts Museum, Dordrecht. In the public domain under Creative Commons CC, see 
\mbox{commons.wikimedia.org/wiki/} File:Jan\_Pieter\_Veth.jpg.}
\label{fig:self}
\end{figure}

Jan Veth was born in 1864 in the city of Dordrecht, son of an  ironmonger. On his maternal side, he descended from a well-known Dordrecht family of painters and a lineage of gold- and silversmiths and from his paternal side of copper- and blacksmiths. He attended the local HBS (Higher Citizen’s School), a type of secondary school instituted in 1863 by the founder of liberalism in Dutch politics and then Prime Minister Thorbecke to provide education preparing boys for a career in commerce and industry rather then academia. Instead of the Greek and Latin of the Gymnasium it provided more extensive studies in mathematics and natural sciences and modern languages (English, French and German). As often told (Willink, 1981), because science was taught by PhDs, one of its major, but unintended effects was for it to become eventually the primary route to university science studies and the cause of the relatively large proportion of Dutch Nobel Prizes early in the twentieth century. The Veth family being part of the middle class made it quite natural for Jan Veth to enter the HBS for secondary education. During his HBS schooling Veth  was noted for his drawing talents. His drawing teacher Adrianus Jacobus Terwen (1841--1918) gave him extra lessons so that he qualified for the Rijksacademie van Beeldende Kunsten (National Academy of Fine Arts) in Amsterdam, for which he passed the entrance exam and enrolled in 1880. 

With some of his fellow students, he was the founder of the artist association {\it Saint Lucas} ‘for development in all subjects taught at the Academy, and to promote friendship among [its] students’, as the founding declaration proclaims. The patron saint of painters, St. Luke, is the namesake.  The association was recognized by Royal decree in 1882 and is still active. Although initially meant as a student society, Saint Lucas was transformed into a general artists' society in 1887. From 1908 to 1948, Queen Wilhelmina was patron of the society. Another initiative Veth was heavily involved in was {\it The Nederlandsche Etsclub} (Netherlands Club of Etching), which was founded in 1885. It was an association of graphic artists for which the  purpose was to bring the etching technique back into the public awareness. It dissolved itself in 1896. Fig.~\ref{fig:self} shows Veth in this period in a self-portrait.

Veth married in 1888 to Anna Dorothea Dirks (1863--1929). They had five children, which I list here since they appeared in the letters cited below with their nicknames, which were apparently used extensively, between square brackets:  Saskia [Kik] (1889--1969), Alida Johanna [Ila] (1891--1960), Gerda [Polle] (1894--1966), Anna Cornelia (1895--1896), and Justus [Joost] (1897--1942). Anna Cornelia died while still a baby and has no nickname. 

After this marriage the couple settled in the city of Bussum in the district het Gooi (often written as ‘t Gooi),  some twenty kilometers to the south-east of Amsterdam around the city of Hilversum. The name derives from ‘gouw’, a region, shire or district.  It also encompasses the village Laren, which together with Bussum developed (some would say revived) around that time into a center of arts. In those periods Veth, who himself also wrote poetry, became intimately acquainted with the movement of the {\it Tachtigers} (Eightiers), a group of writers and poets, who developed a new approach to literature. Veth the poet is seen as part of this movement. He became good friends with  leading members, such as   Frederik Willem van Eeden (1860--1932), for whom he designed a cover for his well-known novel {\it De kleine Johannes} and Albert Verweij  (1865--1937), of whom he produced a painting that would become one of his best known and most praised ones.  He published his poems mostly in literary periodicals, particularly well-known and influential ones as  {\it De Gids},  that had appeared since 1837, and later {\it De Nieuwe Gids}, but also as a separate publication in 1920, entitled  {\it De zwerver spreekt en andere gedichten} (The wanderer speaks and other poems).

\begin{figure}[t]
%\sidecaption[t]
\begin{center}
\includegraphics[width=0.98\textwidth]{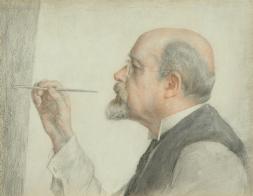}
\end{center}
\caption{\normalsize Jan Veth by Wilhelm Christian `Georg’ Rueter (1875--1966). Undated, from Veth’s appearance it must have been produced some time around him producing the paintings discussed in this paper. Courtesy Simonis \&\ Buunk, art dealers. Credit line: `W.C. `Georg' Rueter (1875-1966), \textit{Prof. Jan Veth behind his easel,} colored pencil on paper, $27.1\times 33.6$ cm. Private collection; formerly collection Simonis \&\ Buunk’.}
\label{fig:Rueter}
\end{figure}

While originally a painter of landscapes Veth quickly developed into a maker of portraits. In order to generate commissions he needed to become known by a wider audience. Around 1890 he turned himself into a lithographer. He had taken up the plan to publish a series of lithograph portraits of well-known contemporaries. Lithography was a process well suited for mass reproduction and on top of that relatively cheap. He was able to convince the editor and publisher of the weekly magazine {\it De Amsterdammer} to publish such a series of portraits. This  magazine was founded in 1877 as a weekly for commerce, industry and art.  One condition was that the litho was good enough to produce 5000 to 6000 copies. The magazine  became popularly referred to as {\it De Groene Amsterdammer} (the Green Amsterdammer, on the first page green ink was used), because a daily newspaper {\it Amsterdammer} had appeared, starting in 1883. Much later, in 1935, this became the official name. The weekly still exists as an online magazine.

In this magazine and later also in the social-cultural and literary weekly {\it De Kroniek} (the Chronicle), which has been published from 1895 to 1907,  altogether more than fifty lithographic portraits made by Veth appeared of well-known Dutch persons. With the publication of these lithographs, he became known to a wider audience as an able and leading portraitist. His name established, he started receiving  commissions from home and abroad. He regularly visited Germany, the U.K. and twice the U.S.A. to work on portraits. His portraits of professors and other important persons are widely regarded as among the best ever produced. Fig.~\ref{fig:Rueter} shows Veth, as he looked at the time he was involved in painting Groningen professors, the subject of this paper.

But in addition to painting and writing poetry, Veth published articles and books, particularly in the areas of art history and art criticism. For example he wrote about his contemporaries such as Jozef Isra\"els (1824--1911), well-known landscape painter, and leading member of {\it De Haagse School} (The Hague School). But also a book about Rembrandt van Rijn (1606--1669) as part of the preparation for the Rembrandt commemoration in 1906. On that occasion the University of Amsterdam bestowed an honorary doctorate upon him together with a few other Rembrandt connoisseurs. He was also active in organizations concerning conservation of monumental buildings and their interiors, etc. In 1917 he was appointed extraordinary professor at the National Academy of Fine Arts in Amsterdam. In 1921-1922 he made a trip to the Dutch East Indies, which resulted in a number of sketches of landscapes and a beautiful painting of the Borobudur. In 1923 he became a member of the Netherlands Royal Academy of Arts and Sciences.

In 1925 problems with his gallbladder that had been manifest for some time (see below) caused Veth to resign from a number of functions, including his professorship. Surgery was in order, but the operation did not produce the intended improvement and he died soon after this at age 61.

\section{Kapteyn's forty years professorship in 1918}

In 1918, Kapteyn celebrated his 40th anniversary as professor at the University of Groningen. For this occasion his portrait was painted by Jan Veth. In the Kapteyn Room of what is now the Kapteyn Astronomical Institute an album is kept that must have been presented to Kapteyn and his wife on this occasion (see Fig.~\ref{fig:Album1}). The first page  says:
\begmarg
On the 20th of February 1918, when it was 40 years after Jacobus Cornelius Kapteyn took up the professorship in astronomy at the University at Groningen, his friends and students have presented his portrait, painted by Dr. Jan Veth, to Mrs. C.E. Kapteyn-Kalshoven.
\endmarg
The album shows on small pieces of cardboard paper, pasted into the album behind a passe-partout frame,  the signatures of all contributors, in alphabetic order. The first two pages, however, had been reserved for the signatures of a few persons who were very special to him (see Fig.~\ref{fig:Album2}). Here I identify these, while very briefly summarizing some aspects of Kapteyn’s career.

First there were colleague astronomers that meant most for  Kapteyn’s career and had grown to be special friends: Mrs. Isobel S. Gill, widow of Sir David Gill (1843–1914), director of the Royal Observatory at Cape of Good Hope, George Hale, as we have seen director of the Mount Wilson Observatory near Pasadena, California, Edward Charles Pickering (1846--1919), director of Harvard College Observatory, Cambridge, Massachusetts and very important contributor to the {\it Plan of Selected Areas}, Anders Severin Donner (1854–1938), director of Helsingfors Observatory at Helsinki and longtime collaborator and supplier of photographic material, Karl Friedrich K\"ustner (1856–1936) also providing photographic material  as director of the Bonner Sternwarte and Robert Thorburn Ayton Innes (1861–1933), successor of David Gill. They all were strong supporters of the {\it Plan of Selected Areas} and important contributors to its progress, Innes having taken over responsibility for the contribution of the Cape Observatory after Gill’s retirement.

\begin{figure}[t]
%\sidecaption[t]
\begin{center}
\includegraphics[width=0.555\textwidth]{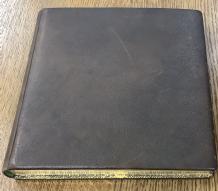}
\includegraphics[width=0.425\textwidth]{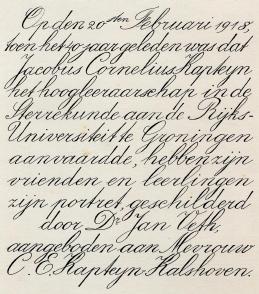}
\end{center}
\caption{\normalsize The album accompanying the presentation of a portrait of Kapteyn to his wife painted by Jan Veth on the occasion of his fortieth anniversary as professor at the University of Groningen. It resides in the Kapteyn Room in the Kapteyn Astronomical Institute. The first page on the right has been translated in the text. Courtesy Kapteyn Astronomical Institute.}
\label{fig:Album1}
\end{figure}

\begin{figure}[t]
%\sidecaption[t]
\begin{center}
\includegraphics[width=0.50\textwidth]{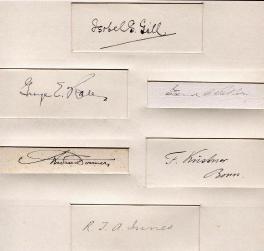}
\includegraphics[width=0.48\textwidth]{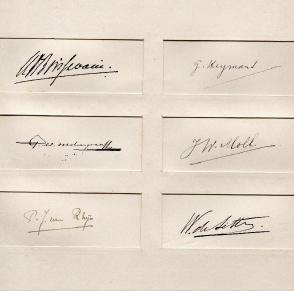}
\end{center}
\caption{\normalsize The first pages with names of contributors of the album accompanying the presentation of a portrait of Kapteyn, painted by Jan Veth on the occasion of his fortieth anniversary as professor at the University of Groningen. The album contains signatures  of those that contributed to the costs. This first two of these pages have the signatures of persons that were special to Kapteyn. On the left Kapteyn's closest colleagues, friends and most important collaborators (Mrs.) Gill, Hale, Pickering, Donner, K\"ustner and Innes, on the right his friends Boissevain, Heymans, Molengraaff and Moll, and his students van Rhijn and de Sitter. Courtesy Kapteyn Astronomical Institute.}
\label{fig:Album2}
\end{figure}

Then there were four of his longtime closest friends:  Ursul Philip Boissevain (1855--1930),  Professor in Ancient History at Groningen, Gerard Heymans (1857--1930),  Professor of Philosophy and Psychology, also at Groningen, Gustaaf Adolf Frederik Molengraaff (1860--1942), Professor of Geology at the Delft Polytechnic School (before 1905 the Delft Institute of Technology and now since 1986 the Technical University of Delft),  and Jan Willem Moll (1851--1933),  Professor of Botany and Plant Physiology. Boissevain and Heymans were very special friends. On most Monday afternoons Kapteyn and these two friends walked from Groningen to the nearby village of Haren, a walk of some 6 km (and back of course). They talked about all kinds of subjects, watched birds and explained to each other how their work was going.
Molengraaff was another special friend indeed; Kapteyn’s second daughter, Henriette Hertzsprung-Kapteyn in 1928 wrote in her biography of her father (Hertzsprung-Kapteyn, 1928; from my translation, p.59);
\begmarg
He asked Prof. Molengraaff, the geologist from Delft, who every year undertook an excursion with his students, if he could join in these trips, and this was gladly granted. So, a few times he went along to far away places as if he were the youngest and most enthusiastic of students.
\endmarg

Finally, Moll had been professor in Groningen since 1890 (Kapteyn since 1878), while both were born in the same year. Kapteyn had spent quite an effort making the biologists familiar with statistics. I will return to this below in some more detail.

\begin{figure}[t]
%\sidecaption[t]
\begin{center}
\includegraphics[width=0.98\textwidth]{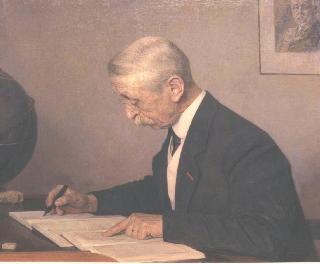}
\end{center}
\caption{\normalsize Painting by Veth of Kapteyn working behind his desk. It resides in the Kapteyn Room in the Kapteyn Astronomical Institute of the University of Groningen. The painting has been produced in 1917. Oil on canvas, $74.5\times 84.5$ cm. Courtesy Kapteyn Astronomical Institute.}
\label{fig:JCKdesk}
\end{figure}

Finally  there were his two most prominent students up till then (Jan Hendrik Oort is the third in this category, but in 1918 he was only a first-year student of astronomy, physics and mathematics in Groningen).  Pieter Johannes van Rhijn  (1886--1960) succeeded Kapteyn as director of the Astronomical Laboratory in Groningen upon the latter’s retirement in 1921 (on that occasion named after Kapteyn), and dedicated his career to continuing the research program of his famous teacher, among which overseeing the completion of the {\it Plan of Selected Areas}. Willem de Sitter (1872--1934) had become director of the Observatory at Leiden; he did fundamental work on the system of Galilean satellites of Jupiter, but is now best known to a wider audience for his work on cosmology and Einstein’s theory of General Relativity, culminating in the model of the Universe that became known as the `Einstein-de Sitter Universe' (Einstein \&\ de Sitter, 1932)
\bigskip

\begin{figure}[t]
%\sidecaption[t]
\begin{center}
\includegraphics[width=0.90\textwidth]{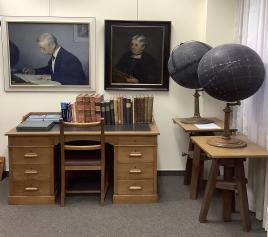}
\end{center}
\caption{\normalsize   Veth’s painting at its present location in the Kapteyn Room in the Kapteyn Astronomical Institute of the University of Groningen. Below are his desk, which is the same one as in the painting, with his most important publications on top. The box on the left contains the album of signatories contributing to the commission of the painting. The chair is almost certainly not the one Kapteyn used. The painting on the right is his wife, painted in the 1930s, when she had been a widow for more than a decade. This painting is by Lizzy Ansingh, and is a gift to the Kapteyn Astronomical Institute by the Kapteyns' great-granddaughter Wilhelmina Henriette de Zwaan-Kaars Sypesteyn for display in the Kapteyn Room and stipulated by her to be located next to the Kapteyn painting. On the right  Kapteyn’s globes with blackboard coating to use for drawings with chalk, to represent features or check constructions or orientations on the celestial sphere. Courtesy Kapteyn Astronomical Institute.}
\label{fig:Kaptroom}
\end{figure}

Ursul Boissevain delivered a speech on the occasion of the presentation of the painting, which has survived in Museum Boerhaave in Leiden. It is in very clear handwriting, each letter penned separately. I quote the final part (my translation):
\begmarg
And we have eagerly seized the opportunity to show our love and affection. We present you with a portrait: we wished that there would be an image of you for future generations appropriate to your dignity. We have found Dr. Jan Veth willing to fulfill our wishes. You can be assured that among those who honor you this way – and among them you will also find the most precious of your foreign friends – there were none who did not full-heartedly join the effort to make this possible. We have only asked your friends to take part in this; I should say: were allowed to take part in this, knowing that you would not have wanted it any other way.

And now, dear Kapteyn, accept this as it is offered to you, also a proof of our true affection. May you be blessed with many years of unrelenting energy to investigate and come to greater understanding, by penetrating more and more deeply into the immeasurable spaces of the boundless universe, the sight of which alone fills the simplest of hearts with respect Then also the wish will be fulfilled that we cherish for your dear and loyal spouse, for your children and grandchildren, and for ourselves, for whom your friendship belongs to the most highly valued among our possessions.
\endmarg

Although the Album states clearly that the painting was offered to Mrs. Kapteyn, the speech is worded as if it was offered to Kapteyn himself. I will refer to this matter as if it were the intention to offer it to Kapteyn’s wife.

The painting that is believed to have been offered is shown in Fig.~\ref{fig:JCKdesk}. The person in the top-right is David Gill. The desk and a globe covered with blackboard coating so that one can draw with chalk on it, are indeed  from Kapteyn’s office and now reside in the Kapteyn Room (there are actually two such globes). Fig.~\ref{fig:Kaptroom} shows the painting at its present location in the Kapteyn Room in the Kapteyn Astronomical Institute. The painting to the right is of his wife, Catherina Elisabeth (Elise) Kapteyn–Kalshoven (1855--1945). It is a gift to the Kapteyn Astronomical Institute by Kapteyn’s and his wife’s great-granddaughter Wilhelmina Henriette de Zwaan-Kaars Sypesteyn under the condition that it be displayed next to Veth’s painting of Kapteyn at his desk. The book she holds in her hands is a copy of the biography \textit{J.C. Kapteyn; Zijn leven en werken} by their daughter  Henriette Hertzsprung–Kapteyn (1928). This portrait has been painted in the 1930s by Lizzy Ansingh (1875–1959), who  belonged to a group of female post-impressionist painters  called ‘the Amsterdamse Joffers’. A ‘joffer’ is best translated as ‘missus’; these women painters, usually from well-to-do backgrounds, did not rely on painting to support themselves, and promoted acceptance of female artists. Kapteyn’s globes are shown on the right.

On September 16, 1918, during the ceremony of the opening of the Academic Year 1918-1919, the Rector Magnificus referred to Kapteyn’s jubilee  in his summary of the {\it Lotgevallen} (Fates) experienced by the University in the previous year. His words are recorded in the Yearbook of the University of Groningen for the Academic year 1917-1918 as follows: (University of Groningen, 1918, p.29; my translation):
\begmarg
Among the events, which further concern the members of the Senate and deserve mention here, a rare anniversary of special character stands out, the fortieth year of the professorship of our Kapteyn. How his scientific work brought world fame to his name and made the astronomical laboratory of this university world famous; how significant and how rich in influence his education aimed at higher levels of thinking and working was for his students; how wise and beneficial his personal contact, of a natural simplicity that remained the same in spite of all honors, and appreciated by his colleagues and friends; all of this and more was expressed on February 20, 1918 by a series of speeches on behalf of the Board of Trustees, Senate, Faculty, alumni, students, friends and corporations of female and male students at a tribute meeting in this auditorium, at which his portrait painted by Jan Veth was presented as a proof of veneration and affection among a wide circle of persons. This meeting, which was also attended by many interested persons  from elsewhere and where only due to the circumstances of the timing representatives from abroad had to be absent, will remain in the memory of all of us affectionate participants, not in the least because of the striking way in which Kapteyn himself expressed his thanks with a review of his life. He and his desendants will excuse me if I express here the quiet hope which is cherished in our academic community, that in due course the University may become owner of this likeness of Kapteyn, so that for future generations of teachers and students who will work there, a visible memory of one of the greatest scholars who worked here will be preserved. The fact that the portrait is a living work of art, and therefore has its own design and dimensions, as a result of which it differs somewhat from what we are accustomed to seeing in the Senate Chamber, will certainly not be an obstacle to its acceptance and placement. 
\endmarg 
The painting was completely out of tune with the gallery of paintings of professors in the Senate Chamber, which had a fixed size (with minor deviations) and depicted the person's head and shoulders, wearing  academic attire.  The `different design’ here undoubtedly refers to the fact that it is not a portrait as usual, but shows Kapteyn sitting and working at his desk.The `own dimensions' are also anomalous, because it is `landscape’ and the gallery contains only `portrait' formats. Paintings differing from these prescriptions of academic dress of course are in the University’s possession, but these are (with only two exceptions) displayed in other rooms in the Academy Building or elsewhere in university premises. In any case, the painting referred to and presented in 1918 without doubt is the one of Fig.~\ref{fig:JCKdesk}.

\section{Three paintings?}

Before continuing I need to say a few words on this collection of paintings in the Senate Chamber, that the Rector Magnificus referred to. The history of the portrait gallery of Groningen University has been described in much detail in a publication entirely devoted to it, {\it In vol Ornaat} (In full Regalia), by  Oosterheert (2009). Collecting paintings by universities of (former) professors associated with it was a tradition that originated in Germany. In the Netherlands, the University of Leiden was the first to obtain a painted portrait of a famous scholar, in 1596, of Desiderius Erasmus Roterodamus (c.1466--1536), but of course Erasmus never had been associated with this university (which in fact was founded in 1576 well after Erasmus’ death). The first beginning of a collection of paintings of a university’s own professors dates back to 1618, when the University of Groningen, four years after its foundation, had opened its Academy Building, the center of the University, with a Senate Chamber. This room was used by the Senate, the collection of its professors, at that time six in number,to meet and discuss things and decide on matters. The walls of this small room were decorated by paintings of the first four Rectores Magnifici, who  had been in office each for one year. These were produced for this occasion and purpose. 

Before 1850, when  the second, much larger Academy Building was built, there was no organized tradition for extending the collection, and few paintings had been added to it in this period in spite of this early start.  In 1851 the Senate decided to use the additional space in the new Senate Chamber to decorate it with a gallery of its professors and adopted a set of rules to which such portraits should conform, including size and the condition that the person  portrayed should wear an academic gown and jabot (bavette), sometiems with a with bow-tie, and if preferred a beret, and if applicable distinctions such as Royal decorations. The image of the professor should be approximately life size (it showed only head and part of the torso) and it had to be an oil painting. Until then the paintings had been spread over various academic buildings, particularly the library. But now a systematic collection and gallery was started in the Senate Chamber, which included both tracking down old portraits and having new ones made. From then on it grew steadily.

\begin{figure}[t]
%\sidecaption[t]
\begin{center}
\includegraphics[width=0.98\textwidth]{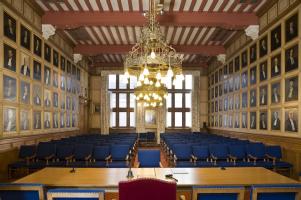}
\end{center}
\caption{\normalsize  View of part of the Senate Chamber in the Academy Building of the University of Groningen. Kapteyn’s portrait is on the right, second row from the bottom, fifth from the right of the picture. In this setting newly appointed professors up to this day are being installed member of the Senate before delivering their inaugural lecture. Having become a member they are asked to be seated in the chair at the table with the Rector Magnificus, Dean of the repective Faculty and a few other high officials seated in the chairs in the front.  In the public domain in Wikimedia: commons.wikimedia.org/wiki/File:Senaatskamer\_in\_de\_oostvleugel\_\linebreak[4]
van\_het\_academiegebouw\_op\_de\_eerste\_verdieping\_-\_Groningen\_-\_20416188\_-\_RCE.jpg.}
\label{fig:gallery}
\end{figure}

The collection was almost lost in the beginning of the twentieth century, when it was decided that a major refurbishment of the Academy Building was in order. This had been almost accomplished over a number of years with in 1906 still the outer walls remaining to be treated. That is when disaster struck: removal of old paint was done using paint burners, which caused a major fire, destroying the entire building. Fortunately all paintings could be removed in time, including the four original ones. It did not take very long to rebuild, and a new Academy Building was inaugurated in 1909. It had a further expanded Senate Chamber where the gallery has since been  displayed on the walls, (see Fig.~\ref{fig:gallery}).

There are only two paintings of professors not wearing academic attire (we will meet one further on in this paper). An obvious person conspicuously missing for a long time was Fritz Zernike (1888–1966), winner of the 1953 Nobel Prize for Physics (who as a student had worked for some time as Kapteyn’s assistant). A portrait of him did and does decorate a wall in another room in the Academy Building, but on that he was not wearing a gown and so it did not qualify for addition to the Senate Chamber gallery.  This situation was repaired in 2003 when a painting of him according to the rules was especially produced. 

The gallery of portraits of professors for a University is a statement of its respectability,  importance and historical significance. Professors sometimes themselves, but usually their descendants, family, students and admirers, bring together the necessary funds and offer the portrait to the university. Some kind  of judgment of the significance of the professor involved is required for acceptance, but that does not lead in practice to serious disagreements. Later in the twentieth century the willingness to offer such paintings had decreased somewhat, while the Governing Board actively discouraged the practice when the walls of the Senate Chamber were completely covered. After the year 2000 the walls of the Faculty Chambers were formally designated to receive such decorations and the tradition revived and willingness to offer new paintings increased again. It still is very much alive (it currently consists of well over three hundred paintings and each year of order ten are added) and will probably be alive for some time, at least as long as there are rooms in the Academy Building that can accommodate further additions. 
\bigskip

 Adriaan Blaauw (2000) told an interesting story in his chapter in the {\it Legacy Symposium} (van der Kruit \& van Berkel, 2000; p.4):
\begmarg
I was told by the late Pieter J. van Rhijn, who was Kapteyn’s close collaborator and successor and my predecessor, that Veth was inspired to paint Kapteyn the way we see him here, by a remark made by Mrs. Kapteyn. She felt little sympathy for [the original] version, also made by Veth and donated to Kapteyn by friends and colleagues of Kapteyn, which shows Kapteyn posing for the painter. ‘This is not how I am used to seeing my husband’, she said, as van Rhijn conveyed to me. The way she did see him – at work at his desk – is depicted by the portrait in the Kapteyn Room. In the upper right corner of the painting, Veth sketched David Gill, the close collaborator and a friend of the Kapteyn family. The painting was acquired by the family and donated by Kapteyn’s heirs to the University of Groningen around the year 1960, to be placed in the Kapteyn Laboratory. The donation was the result of an approach, initially by van Rhijn in September 1957, to Kapteyn’s heirs, in particular to his daughter Mrs. Noordenbos–Kapteyn (widow of the Amsterdam professor of surgery W. Noordenbos [this is Kapteyn’s elder daughter Jacoba Cornelia]), who at that time lived in England near her daughter Maria Newton–Noordenbos. After consulting Mrs. Noordenbos–Kapteyn and the children, it was decided that the painting would be donated to the Laboratory after her decease. A lucky circumstance, which may well have facilitated the transfer, was the fact that the late Maria Newton–Noordenbos, Kapteyn’s grand-daughter, was a class-mate of this author [that is Adriaan Blaauw] in grammar school in Amsterdam in the years 1928–1932. [...]
\endmarg

\begin{figure}[t]
\sidecaption[t]
%\begin{center}
\includegraphics[width=0.64\textwidth]{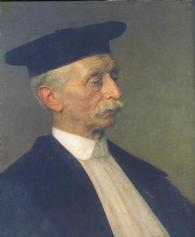}
%\end{center}
\caption{\normalsize The painting of Kapteyn that decorates the Senate Chamber of the University of Groningen. It is signed by Veth in the top-right corner and dated 1921. Oil on canvas, $70\times 55$ cm. Courtesy University of Groningen. }
\label{fig:JCKSenate}
\end{figure}

So there was an earlier version that Mrs. Kapteyn did not like. What happened to it? Also, there is a painting of Kapteyn in the Senate Chamber in the Academy building (see Fig~\ref{fig:JCKSenate}), also painted by Jan Veth. Are there three paintings of Kapteyn?  I continue with Blaauw’s narrative.
\begmarg
But what became of the 1918 painting donated by friends and colleagues? The walls of the Senate  of Groningen University are covered with a mosaic of paintings of retired professors. Among them, somewhere near the center of the west wall, we see the one of Kapteyn. According to the rules set by the University for such portraits, it shows Kapteyn dressed in his University gown and cap. It is signed by Jan Veth and carries the year 1921, i.e., that of Kapteyn’s retirement. This raises the question: did Veth paint Kapteyn again in 1921, three years after he produced the two paintings mentioned before? The question has puzzled historians – for if indeed Veth did so, where then is the 1918 painting?  It is nowhere referred to among Veth's descendants who are known to guard so preciously the whereabouts of what reminds them of their famous ancestor, and also van Rhijn never referred to it. The most natural solution seems to be that, when the time came for delivering a ‘retirement portrait’, the 1918 painting was adapted by Veth himself to the University’s special conditions: he adjusted Kapteyn in the way prescribed. A close inspection of the painting performed in February of the year 1992, in the presence of the curator of the University Museum, Mr. F.R.H. Smit, supported this supposition: traces of Kapteyn’s head of hair seem to betray Veth’s disguising efforts.
\endmarg

\begin{figure}[t]
\sidecaption[t]
%\begin{center}
\includegraphics[width=0.64\textwidth]{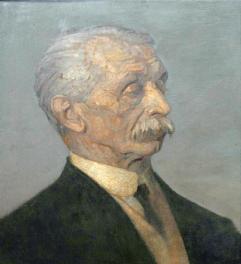}
%\end{center}
\caption{\normalsize Reproduction of a painting of Kapteyn, produced by Jan Veth as a preliminary design of his first painting of 1917. It is in the possession of Kapteyn’s greatgrandson Jacobus Cornelius (`Jack’) Kapteyn, apparently originally donated by Veth to George Reuter, but returned to the Veth family after Rueter’s decease and subsequently passed on to the Kapteyn family. Oil on wood, $40\times 48$ cm. Courtesy Jack Kapteyn. }
\label{fig:JCKJack}
\end{figure}

It is important to stress that if Blaauw’s hypothesis is incorrect there has to have been a third painting of which there is no trace left. The usual remark made for this case is that this painting, produced in 1917 before Veth started painting Kapteyn behind is desk, is lost. This is a very unsatisfactory explanation, because Veth paintings do not easily get lost without any reasonable explanation or realistic sequence of events.
\bigskip

The presentation of my scientific biography {\it Jacobus Cornelius Kapteyn: Born investigator of the Heavens} (van der Kruit, 2015)  in January 2015, was attended by Jack Kapteyn (Jacobus Cornelius!), grandson of Kapteyn’s son Gerrit Jacobus. He had a painting with him to be displayed during the proceedings, that also was painted by Jan Veth (see Fig~\ref{fig:JCKJack}). It is painted on a wooden panel, but not signed or dated. A note glued to the back (Fig.~\ref{fig:JCKJackback}) reads:
\begmarg
Portrait Prof. J.C. Kapteyn [illegible] bestowed to Georg Rueter. In 1967 when clearing out his studie returned to the Veth family.
\endmarg
Rueter is the person who made the drawing of Veth in Fig.~\ref{fig:Rueter}. Maybe Veth gave it to Rueter in appreciation of this. It ended up in an unknown manner with a Jaap Kapteyn, a cousin of Jack, who passed it on to the latter. The Veth descendants must have felt it appropriate to donate it to the Kapteyn family.    

This is clearly a preliminary study. The facial expression and posture, as well as the angle from which he is seen, are strikingly the same as in the painting with the academic gown; Kapteyn is wearing the same clothes as behind his desk in the other painting. The conclusion would be that this is a a preliminary study of the original painting Veth produced. Then this shows what this painting looked like. The painting Veth produced subsequently of Kapteyn behind his desk shows him in the same clothes, because Kapteyn would have had to pose again for Veth and quite naturally would have chosen the same neat clothes, possibly especially acquired for the occassion,
\bigskip

\begin{figure}[t]
\sidecaption[t]
%\begin{center}
\includegraphics[width=0.50\textwidth]{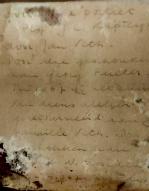}
%\end{center}
\caption{\normalsize Note pasted to the back of the wooden board on which the preliminary study in Fig.\ref{fig:JCKJack} of a Kapteyn painting  had been painted by Jan Veth. See text for transcription and translation of the note. Provided and courtesy by J.C (Jack) Kapteyn, owner of the painting.}
\label{fig:JCKJackback}
\end{figure}

I note that there is no doubt that indeed it was Veth who produced this preliminary version. However, the question arises, why Veth produced this. Was it common practice for him? If he did this on a regular basis, what did he do with them? And if not, why did he do this in Kapteyn’s case? I will encounter a possible second case of this practice below (see end of section 6).
\bigskip

It is of interest to also have a look at what Veth's biographer Huizinga (1927) had to say about this. Johan Huizinga (1872--1945) was a famous historian, particularly known for his monumental work {\it Herfsttij der Middeleeuwen} (Autumn of the Middle Ages), published in 1919. He was the son of Groningen Professor Dirk Huizinga (1840--1903), a Professor of Physiology. The elder Huizinga had been a very good friend of Kapteyn. He had allocated a few rooms in his Physiological Laboratory when Kapteyn set out to measure the plates from Cape Town for the {\it Cape Photographic Durchmusterung}. Actually when under Huizinga's successor Hartog Jacob Hamburger (1859--1924) the laboratory moved to a new, larger location in 1911, this building behind the Academy Building had been assigned to Kapteyn to house his Astronomical Laboratory. Johan Huizinga had studied history in Groningen, obtaining his PhD in 1897. In 1905 he had been appointed Professor of General and Dutch History in Groningen, until he moved to Leiden in 1915, where he became a Professor of History. Huizinga the historian was well acquainted with Kapteyn; in fact he and his friend Willem de Sitter (1872--1932), the first student to obtain a PhD under Kapteyn and Professor of Astronomy in Leiden, in 1925 would take up the plan to write a biography of Kapteyn (see Preface in van der Kruit. 2015).

We know this because Willem de Sitter wrote a short ‘letter to the editor’ in the section ‘correspondence’ of the journal {\it The Observatory}  (de Sitter, 1925), which read in part as follows: 
\begmarg
My friend J. Huizinga, Professor of History in this University and myself intend as joint authors to write a biography of Prof. J.C. Kapteyn. [...]
We shall be greatly indebted to any friends of Kapteyn, or other astronomers, who may be willing to assist by communicating to us any detail they may happen to know, or any important point of view or material they may have at their disposal, regarding Kapteyn’s many national and international connections with astronomers and institutions all over the world, or his personal relations with his many friends and acquaintances, or any information, even of anecdotal character, which may help us to make the picture of the man as complete as possible. Letters written by Kapteyn will, of course, be most valuable.
\endmarg

Nothing came of this, and we do not know why. Now in his biography of Veth (Huizinga, 1927), Huizinga wrote (p.83; my translation):
\begmarg
Veth liked to paint a model, that interested him, more than once. […] He did it with Kapteyn and Kuenen.

	J.C. Kapteyn, the astronomer at Groningen, was a man after Veth's heart. In 1917 he finished his portrait. ‘Professor Kapteyn is one of the most brilliant Dutchmen I have met. A man of world renown with the simplicity of a simple ship's captain, and thereby of incredible vivacity and clarity of mind.’ When the first proof was ready, a colleague's wife came to see it, and said in all innocence, that the portrait really looked more like Prof. Kapteyn than .... Kapteyn himself, who was sitting next to it.

[Huizinga then quotes Veth as follows:] `I don't believe this is a bad painting, but tomorrow I will start another one of the same remarkable man. He usually spends five or six hours a day working on mathematical tables, and that's how I see him sitting in front of his window when I approach to enter the laboratory. Well, that's how I wanted to paint him also a second time, because I think I could make something very convincing that way. For he is really sitting at that table in concentrated self-oblivion.’
\endmarg

Unfortunately that is all Huizinga wrote about Kapteyn. Interestingly — and importantly for this discussion — , in his {\it  Preliminary catalog of the painted and graphic works of Jan Veth} at the end of the biography, Huizinga lists two paintings of Kapteyn, both in the year 1918 as the following entries (p.236/7; my translation):
\bigskip

848 Prof. Dr. J.C. Kapteyn, bust, oil, Senate Chamber Groningen

849 Prof. Dr. J.C. Kapteyn, half-length, oil, Mrs. Kapteyn-Kalshoven, Hilversum
\bigskip

\noindent The final entry in each line is the location at the time of Huizinga’s writing of the book and preparing the list.
\bigskip

The `half-length’ is my translation of `kniestuk’ (literary knee piece), which refers to a portrait, in which the person is depicted intermediate between bust and full-length. According to Huizinga the painting in the Senate Chamber has been painted in 1918(!) and even precedes the one with Kapteyn at his desk. This would fit prefectly with Blaauw’s hypothesis. There is no entry for 1921, when Veth would have produced a separate one for the Senate Chamber. This is also what would be expected if the Blaauw hypothesis were  correct. Huizinga does not tell in detail on what he based all of this, but even then it is strong support of the notion that the painting in the Senate Chamber actually is the one produced first in 1918.

That Huizinga could speak of all this with some authority is evident from the following citation (Huizinga, 1927, p.81; my translation): 
\begmarg 
It was also the interesting man, provided he had a striking face, whom he wanted to portray. He looked for him everywhere, and with a certain preference for one having an active or scientific life. He wanted workers and doers.
	
It is not a preference for the professors, nor for the letter K, which, to be more specific for a few cases in which Veth pursued important compatriots with pen and brush, that determined the choice of the foursome Kuyper, Kern, Kuenen and Kapteyn. It is, even though it may definitely be called an illustrious group, mainly because this author has known three of them well and has been able to observe the creation of the portraits to some extent, in addition because all four pieces represent Veth's art from different sides at its best. 
\endmarg

Huizinga’s father had been a very close friend of Kapteyn and Huizinga himself had been a professor in Groningen for a decade, so undoubtedly Kapteyn will be one of those three. There is no doubt that Huizinga was very well informed on most of not all aspects of the creation of the Kapteyn paintings.
\bigskip

In the {\it Lotgevallen} of the university in the academic year 1920-1921, the Rector Magnificus reported (University of Groningen, 1921, p.37/8; my translation):

\begmarg
Two portraits were offered and gratefully accepted by the Senate this year; […]; the other of colleague Kapteyn, painted by Jan Veth, offered by his students and to which a place will be assigned by the Trustees in the Senate Chamber. To the givers we offer our warmest thanks.
\endmarg

This suggests Veth producing a painting in 1921 with academic attire, especially for the Senate Chamber.  It apparently was commissioned by his students, who would at least for the larger part already would have donated to the 1918-painting. 

If the painting in the Senate Chamber is {\it not} the original 1917 one overpainted with academic attire, there have to be {\it three} Kapteyn paintings and with his intimate knowledge and information Huizinga would indeed have listed three paintings. We may safely conclude that there is no lost third painting and the one in the Senate Chamber has to be the first one of 1917. The preliminary painting shows that without doubt Veth has not started painting Kapteyn in academic attire. That would be ruled out anyway since that painting was produced to be offered to Mrs. Kapteyn and who would think of portraying him then in academic dress.    
\bigskip

\section{Veth’s letters May to July 1917}

These remarks by Huizinga provide strong support of Blaauw’s hypothesis. One final source appeared only recently in connection with a large exposition by the Dordrechts Museum, referred to above, of Veth’s works in 2023, where many paintings, drawings etc. were displayed. Since the painting at his desk from the Kapteyn Room was displayed there  and the Kapteyn Astronomical Institute was invited to send a representative to attend to the opening of this exhibition, for which my wife and I were chosen. The two paintings of Kapteyn in Figs.~\ref{fig:JCKdesk} and \ref{fig:JCKSenate} were displayed prominently next to each other and some accompanying text quoted from letters of Veth. These descriptions at the exhibition for its visitors in addition to these quotes stated {\it as a fact} that a third painting by Veth has been produced and now is lost.

As it turned out these were letters Veth wrote to his wife when working away from home. Through some correspondence afterwards with the staff of the museum, particularly the curator 19th Century,  Mrs. Quirine van der Meer Mohr, I learned that these letters, residing in the `Regionaal Archief Dordrecht', had recently been made available in electronic form on their Website (Regionaal Archief Dordrecht, 2023).

In all, there are thirteen letters in which, according to the search function, the name Kapteyn appears. Two are from his wife to him, of which in one (from 1906) a teacher Kapteyn of Johan (daughter  Alida Johanna, born 1891) is mentioned. There is also a letter from 1906 from `Kik' (daughter Saskia, born 1889), in which a Mr. and Mrs. Kapteyn appear, who came to look at a bookcase that maybe the Veths had up for sale. These letters are about different Kapteyns. The other letter from his wife Anna Dirks is from 1917, the rest are from Veth to his wife, seven from 1917 and three from 1921. All of these ten letters have remarks concerning paintings of Kapteyn. The letters written by Jan Veth are also present in transcript in typed form (which makes a total of 21 letters in the search function). It is then straightforward to process these letters with a OCR (optical character recognition) application.

I will proceed to go though these letters, and in order to preserve the flavor quoting them extensively, leaving out only parts that are totally irrelevant here. The first one is from Veth to his wife, written in Groningen. He stayed with Jan Willem Moll, Professor of Botany and Plant Physiology, whom we met above. The other persons mentioned will be introduced after the letter. The transcript presented next is my translation into English.

\begmarg
\hfill{Groningen, Thursday evening 24 May '17}

This is a very different atmosphere here than in Rotterdam. Special people and people with ideas. Mr. Moll is wise, reflective and interested in everything. Mrs. Moll is a Mrs. Bastert of many times more intelligence and very well-read. Prof. Kapteyn is one of the most brilliant Dutchmen I have met, a man of a world name with the simplicity of a simple ship's captain, and thereby of an incredible vivacity and clarity of mind. I also met Professor Heymans several times, the philosopher, with whom, naturally, one comes to a profoundly philosophical conversation. Last night I paid him a long visit.

Through Prof. Moll I was introduced  to a female curator in his laboratory who is helping me with an investigation into the virtue of the illustrations in the Cruydt-boeck van Dodonaeus, about which I want to write something for ‘De Gids’.  And I learned a lot from this humble little miss.

So this is all worthwhile. It is only a pity that Groningen is so far from Bussum, and that you don't see any of it. You would also find the Hortus, in which Moll's house is located, beautiful. One can reflect on  the most wonderful flowering plants and the richest formations.

The portraits themselves have only just begun, — that of Kapteyn further advanced  than that of Moll —  and I shall probably have to come here several times more which, at least mentally, will not be boring at all.

But I shall be home late on Saturday evening.

Bye dear Mom
\endmarg

Mrs. Bastert must be some acquaintance of the Veths. Gerardus Heymans we met above as one of Kapteyn's best friends on the first page of the Album associated with the painting for which Veth was in Groningen. He was one of the most prominent philosophers of his time and the founder of psychology in the Netherlands. One of the most famous herb books, illustrating plants and herbs, was the {\it Cruydt-boeck} by Rembertus Dodonaeus (1517--1585). The first edition appeared in 1554. Dodonaeus, professor in Leiden, was the first to organize the plants no longer alphabetically, but according to physiological characteristics. This curator at the biology department is not described by name. It is tempting to think this may be Jantine Tammes (1871--1943), assistent to Moll, who rose to become the first female professor at the University of Groningen, and the second in the country.  The literary periodical {\it De Gids} we have encountered above.

Moll and his wife must have offered Veth to stay with them and work on both paintings. Their house was apparently located in the {\it Hortus Botanicus} of the University of Groningen, which was located at the time within the city of Groningen, not far from the central university buildings. It had been founded in 1626, not long after the founding in 1614 of the university itself. Moll as  Professor of Botany and Plant Physiology was in charge of the {\it Hortus}. Kapteyn and Moll were very good friends. Moll was on the first page of the album in the Kapteyn Room, described above, and was most likely involved in organizing the celebration of Kapteyn's jubilee. Moll and Kapteyn were the same age (born in 1851), so both were due to retire in 1921. Moll had been appointed professor in 1890 (Kapteyn in 1878), but resigned in 1917 as ordinary professor because of poor health (particularly deteriorating eyesight), and had been appointed extraordinary professor until his retirement. He too was presented with a painted portrait offered by `a limited group of former students and friends’ (see citation below), to which Kapteyn must have belonged. The Rector Magnificus continued in his {\it Lotgevallen} for 1918 after the paragraph above concerning Kapteyn as follows (University of Groningen, 1918, p.30; my translation):

\begmarg
As in the past, when he had reached  twenty-five years as a professor, colleague Moll wished in the previous academic year, when he resigned his professorship, that no public tributes would be paid. This wish had to be respected, albeit very reluctantly by many. But the desire to give him, at least privately, a token of appreciation and friendship was too great and too heartfelt for them to resist. So this winter, in all intimacy on behalf of a limited group of former students and friends, his portrait, also painted by Jan Veth, was offered to him. At the risk of appearing immodest, I would like to make it known that the same hope I have just expressed also applies to this picture. 
\endmarg

These  wishes by the Rector Magnificus have eventually become reality for both paintings.

There appears a difference between the two men.  Moll disliked the idea of any formal or public celebration of the milestone of 25 years professorship, but Kapteyn seemed to have approved this on the occasion of his 40 years in that position. Yet, Kapteyn was a very modest man as well, but may have felt that he could not oppose this in the case of the very special milestone of four decades. Moll and Kapteyn were actually more than simply good friends, and I will briefly explain this special relationship. As I mentioned already, Kapteyn had helped biologists, particularly Moll and his assistent Jantine Tammes, with problems of statistics, particularly skew or log-normal  frequency distributions. For more details see van der Kruit (2015, 2021a). For those not familiar with statistics I note that, although in nature many properties (e.g. the size of humans or of berries) are distributed like the so-called normal distribution, more or less symmetric and Gaussian around a mean, many distributions in nature and particularly in biology are far from symmetric (e.g, the volumes, proportional to cubes of diameters, of berries, which is more relevant than diameters). Kapteyn had given lectures  for the biologists and biology students on properties of skew distributions and had actually published on this and because of this had entered into a harsh controversy with British statistician Karl Pearson (1857--1936).  In this context he had also  built a quincunx, a demonstration device for log-normal distribution functions. This beautiful apparatus, an notable piece of Kapteyn heritage, was for many years kept in the biology department, but at some time has been lost (one for the regular normal distribution has survived), but has recently been rebuild using Kapteyn’s specifications (Lucas \&\ van der Salm, 2017). All in all, Kapteyn had spent so much time on this matter that David Gill had written him on 27 March, 1907\footnote{Gill’s letters to Kapteyn, and of many other documents relating to Kapteyn, are available in electronic form online through my dedicated Kapteyn Website, accompanying my biographies of him (van der Kruit, 2015, 2021a).}:

\begmarg
I am glad to hear that you confess to a temporary possession by an evil spirit. Some form of exorcism is necessary -- and I wish to administer it, if I can. I do think that in astronomy at the present time there is nothing comparable in interest with your work [...]
\endmarg
\bigskip

We learn from this letter from Veth to his wife  that Veth started the painting very well in advance of Kapteyn's professorial anniversary in February of the following year, 1918. The first work on it had already taken place in May 1917. And it was synchronous with a painting of his special friend Jan Willem Moll.
\bigskip

Two and a half weeks later, Veth was back in Groningen. He wrote to his wife (my translation):

\begmarg 
\hfill{Groningen, Monday evening, June 11, '17}

Again I am working very hard, and — curiously —  while the last time the portrait of Prof. Kapteyn seemed to come along much better, this time that of Prof. Moll proceeds more vigorously. The long light gives me long days, because actually I paint pretty much until eight in the evening. Of course not all the time while my model is sitting in front of me, but partly still working on it from memory.

And in the meantime I dabble a bit with Dodonaeus, about whom I now know a little more. But I can't spend much time on that. The painting comes first and that now has got hold of me. But moreover I noticed that I am doing my host and hostess a great favor by reading to them. She has to do it that all the time and she sometimes gets enough of it. This is how a substitute brings her solace. I'm actually reading the Werther now, which is always a very nice thing. Madam knew it, but her husband did not. The `Meister’ would perhaps be better, but it is a bit long and doesn't pick up so quickly.

These evenings I have seen something very bizarre here. In the hortus there is a large cactus greenhouse which in itself is something wonderful. And in it no less than twenty cacti were blooming at the same time the night before last. I can't imagine something so mysterious, and I was so sorry that you could not see it. Last night there were three that were even bigger and each one of them even more beautiful. But those twenty moon-sized stars, shining together
in between all those twisting cactus leaf shapes, that's something I'll never
forget. [...]

Bye dear mom
\endmarg

The moon-sized stars were of course the circles of flowers on top of the cacti. Reading aloud obviously has to do with Moll's declining eyesight. Sitting for a painted portrait in those days meant posing a long, long time, sitting still many hours on end and therefore was quite taxing. This is different now; when a painting of me was made for the collection of the University of Groningen in the Academy Building a large fraction of the work was done from a set of photographs taken by the painter and I actually sat only twice for a couple of hours. 
\bigskip

 Veth was back again a month later (my translation):

\begmarg
\hfill{Groningen July 10, 1917}

Nay, it was not easy this time in Groningen. No really major difficulties, but really no  easy ride either.  In the first place perhaps because, especially in the mornings, I felt all the time a bit dizzy. It's better in the afternoons and evenings and also once I'm at work, it doesn't bother me really. But then I had a misfortune with the portrait of Prof. Moll. He told me that he would leave for Eerbeek on a July 13, and  I had arranged my visit here so that I could have his portrait finished, or nearly finished, before then. But he told me on Saturday, that he needed the last four or five days before his vacation to prepare for it. So I can do no  more work after Sunday. This was a a setback for me. And with Prof. Kapteyn things didn't go quite as I hoped. It is so terribly difficult, this  painting of portraits! I find it more difficult almost every day. Sometimes I think I will not  be able to  do this work, moving around from here to there all the time, for very much longer. But then again I find it so beautiful and worthy that I resolve to defy all difficulties and quietly continue on. Only it frightens me sometimes so much to be away from home, and let  life pass so quickly without catching  my breath.

This afternoon I just about finished Prof. K's portrait. And a professor's wife whom he knows, came to see it, who had not seen it yet. And she was so delighted that, without any `Sch\"ongeisterei’ and in all innocence, said, that my portrait really resembled Prof. Kapteyn much more than ..... the man who sat next to it did, who was none other than my model himself. I wished with all my heart, that I would be convinced myself, -- although I don't believe, that it is a bad painting.

But tomorrow I shall begin another painting of this remarkable man. He usually sits about five or six hours a day at his desk, working on mathematical tables, and that is how I see him sitting behind his window when I am coming from outside going into the laboratory.  Well, that is how I want to paint him again, because I believe I could make something very convincing. For he is sitting at that table in concentrated self-forgottenness. I am now starting a sketch of it, more or less like
that painted sketch of the old Bosman, which is somewhere in my box. In 
September, when I come to finish Moll, I will continue it further.

In the meantime I do not yet know how many days I will stick around here.

[Soms organizatorial matters concerning his family members and others, unrelated to the paintings.]

Bye dear Mom
\endmarg

The sketch Veth announced he would make apparently did not survive. I will comment on the Bosman sketch below.
\bigskip

\begin{figure}[t]
\sidecaption[t]
%\begin{center}
\includegraphics[width=0.64\textwidth]{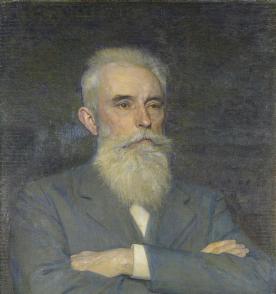}
%\end{center}
\caption{\normalsize Painting of Jan Willem Moll, Professor of Botany and Plant Physiology, produced at the same time as the first painting of Kapteyn for which the preliminary try-out version is in Fig.~\ref{fig:JCKJack}. This painting now is in possession of the University of Groningen and resides in the University Museum. Oil on canvas, $81\times 62$ cm. Courtesy University Museum Groningen.}
\label{fig:Moll}
\end{figure}

There is no further mention of the Moll painting in the letters. It did of course get finished and the result is shown in Fig.~\ref{fig:Moll}.  Former director of the University Museum Franck R.H. Smit wrote to me (my translation):
\begmarg
The portrait Veth made of Moll hangs in the stairwell of the University Museum.  It originally hung in the Botanical Laboratory. We found it in the mid-1990s at the Botanical Museum which  the Faculty [of Mathematics and Natural Sciences] had stored in the Biological Center. It then had a hefty rupture to the right of the face. Culprit unknown, probably happened while moving. We had that repaired.
\endmarg
In fact another portrait of Moll exists, made by Tjitske Maria van Hettinga Tromp (1872--1962) already in 1912 and in the 1960s donated to the University of Groningen, that does indeed decorate the Senate Chamber. Interestingly Moll is not wearing an academic gown there either but a dark suit -- one of only two exceptions in the Senate Chamber that slipped through the cracks (Oosterheert, 2009).

So Moll's portrait was more or less finished in July. It obviously was not meant for the professor’s gallery in the Senate Chamber. It also is a bit large for that, although would probably just fit.

The story Blaauw told suggested that the first portrait was not offered until February of the following year. And that subsequently Mrs. Kapteyn was not satisfied with it (`this is not how I know my husband, I know him working at his desk') and Veth therefore started another painting. So this timeline of Blaauw's story is incorrect. Already in the summer of 1917, Veth decided to make a second painting with Kapteyn at his desk. We should not dismiss Blaauw that easily, since in a sense he had heard it firsthand, because van Rhijn, who told him the story, was an important member of the committee organizing the celebrations and will have been involded in this development. On the other hand, when  writing the Oort biography I have also experienced cases, where Adriaan's memory proved to be not so accurate in the details. The fact that Mrs. Kapteyn did not like the first painting and preferred to see him painted behind his desk, we need not question.

The question now is what prompted Veth to start a new painting of Kapteyn to begin immediately after the first was finished. Only the fascination with the man? And then what would he do with it when finished? Of course there probably was no money with the patrons to commission another painting by Veth. The story of the professor's wife suggests that several others visiting Moll saw the portrait in development. So perhaps Mrs. Kapteyn saw it too (Kapteyn and Moll lived only a short distance from each other) and had already made her objections known at the time. That may then have played a role in Veth's decision to start the new painting, and because doing this appealed to him so much, he would have decided to do it without a fee.

It may also be possible that Veth started the second painting completely on his own initiative for his own satisfaction and pleasure, and then with the intention of keeping that second work himself. So that when it turned out that Mrs. Kapteyn did not like the first, he gave the second painting to her and instead kept the first one for himself. A fact is that in February 1918 the second, and definitely not the first one  was offered.

\begin{figure}[t]
%\sidecaption[t]
\begin{center}
\includegraphics[width=0.80\textwidth]{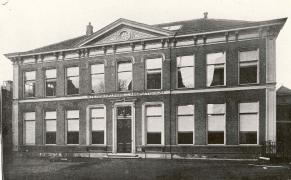}
\includegraphics[width=0.80\textwidth]{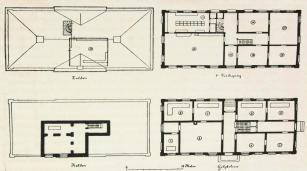}
\end{center}
\caption{\normalsize Top panel. The Astronomical Laboratory seen from the front as one approached it in the days when Kapteyn was still working there (after his retirement it was renamed Kapteyn Astronomical Laboratory and this was written out in full above the entrance). Bottom panel: lay-out of the rooms in the building, clockwise from top-left attic, first floor, ground floor, cellar. Some rooms are identified by their numbers in the text. Top panel courtesy Kapteyn Astronomical Institute (top), and bottom panel from Kapteyn (1914).}
\label{fig:Lab}
\end{figure}

And then there is Kapteyn himself. Veth could well decide to make another painting, but Kapteyn had to be prepared to pose for many hours again, even while  he was probably experiencing this as extremely unpleasant. So there also had to be a very good reason for Kapteyn to agree to it. To please Veth when he announced he was willing to do another painting would be a valid reason, although unlikely to convince Kapteyn (he was experiencing increasing urgency to finish at least a first version of a model of the Sidereal System while his retirement was quickly approaching) to spend his precious time on. It would very likely be acceptable to make the sacrifice, however, if it was because his wife then got a painting she really liked. Whatever the case, there must be more to it than just a wish by Veth to make a second painting of Kapteyn sitting behind his desk.

The description that Veth saw Kapteyn at his desk from outside the Laboratory
gives rise to the following remarks. Fig.~\ref{fig:Lab}, top panel shows the front of the Astronomical Laboratory as seen from the outside by a visitor approaching it. So where was Kapteyn’s office?  In an article for the commemorative book for the 300th anniversary of the University, Kapteyn (194) presented a lay-out of the building and a plan of the rooms on each floor (see Fig.~\ref{fig:Lab}, bottom panel). He wrote (my translation):
\begmarg
The upper floor has the office of the professor (10), of the assistant (9), the library (15) and three rooms for the calculators (12), (13), (14).
\endmarg

Room 10 in Fig.~\ref{fig:Lab} is located behind two windows on the upper floor, the third and fourth from the right. The laboratory was located behind the Academy Building, seen from its front a bit to the left (e.g. Fig.~22 in van der Kruit, 1922a), and walking up to the front you could  perhaps see Kapteyn sitting behind his desk on the first floor  from afar, if it was not too bright outside and Kapteyn's study happened to be  lit up brightly. It was certainly not that obvious to a casual visitor of the premises than maybe this remark by Veth would suggest. 
\bigskip

Anna Veth-Dirks wrote a response to the last letter above on the next day in a long letter with personal information, including that son Joost, now twenty years old and studying chemistry, was going to be employed in  a factory with good prospects for a long term contract. She had a paragraph with a reaction to her husband's letter as follows (my translation):
\begmarg
This afternoon I was very happy to receive your letter, which I have been looking forward to for a long time. I understand your difficulties this week and that it is not a matter for just pulling it out of your hat. And also that sometimes you long for some freedom of action. That is why I am so happy that things with Joost are progressing and coming to clarity. It is a pity that you will not be able to finish the Moll portrait this week, and but that of Kapteyn seems to be going very well and that second one like the Bosman portrait that you have here, seems beautiful to me too; and full of expression.
\endmarg

This very likely was Cornelis Bosman (1830--1911) from Alkmaar, some 40 km to the north of Amsterdam, industrialist and director of a steamboat company. According to the Huizinga inventory Veth had made a drawing (not further specified) of him in 1904 and an oil painting in 1905, and at the same time also an oil painting of his wife  Geertje de Groot (1834--1916). These were at the time of the writing of this biography in the hands of members of the Bosman family in Alkmaar. So, why did Veth in 1917 have a Bosman portrait in his home, while he had painted his portrait in 1905? My guess would be that maybe in this case he did also produce a preliminary version that he kept for himself. This either got lost more recently, or may in fact be the drawing he made in 1904 that he at first had kept for himself and which then ended up after Veth’s demise (and before Huizinga put together his listing)  in the hands of a Bosman family member.  

\section{Correspondence in the fall of 1917}

The next letter is from after the summer and in November of the same year (my translation):.

\begmarg
\hfill{Groningen 17 Nov, 1917}

The journey was very convenient on Thursday and I arrived only slightly late
here. Friday morning we went straight to work, but the fireplace in the room where I was painting smoked terribly and we had to go for a walk. After that we arranged things in such a way that, while I was painting in the front room (the only place where the light is good), the  adjoining back room is being fired up. If the weather stays mild, that's fine. But if a real cold should strike, then we would not be able to bring it to a decent temperature.

In the meantime I am hoping for a tailwind. These two days, even though it turned out to be pitch dark at three o'clock, I have carried the painting much further. And I have now agreed that I will work on Prof. Moll on Sunday, Monday and Tuesday. Then it must be finished and then I will work on the writing portrait of Prof. Kapteyn, on which I hope to continue directly. At his laboratory at least there is a lot of heating.

Yesterday night I attended a debating evening of mainly professors and this evening and tonight tonight [accidentally repeated `tonight'] I am going with Moll again to a more intimate meeting at Prof. Heymans'. So one does spend one's time well here.

[Various maters not relevant here]

Write soon. Lots of love
\endmarg
\bigskip

So it is clear that in November, Veth started working on Kapteyn's painting in which he is sitting  behind the desk,  and that was painted at the Astronomical Laboratory. 
\bigskip

A few days later (my translation):\\
\begmarg 
\hfill{Groningen Nov 23, 1917 Friday evening}

The portrait of Mr. Kapteyn (notwithstanding the very short days) now comes along well. It is a very nice painting, but the sitter's posture  is extremely inconsistent and so it happens that I spend half an hour sometimes staring at my model   without seeing him  in the proper posture.  But because of this circumstance one never comes to getting stuck in the details and keeps working on the whole thing all the time. In  the course of an hour this afternoon I've take  the painting a good deal further by changing the interplay between the coat and the head. The first has become less blue and the shades in the face finer. Tomorrow Mr. K. is away and then I will work on the background, which also requires l a lot of work. Then he will be back on Sunday and Monday, and I wonder if I will not then gradually see the end of this fascinating portrait. I only would want it to be more striking.

In order not always to hang around the Molls, I  sometimes go out in the evening. Two days before yesterday I had dinner at Prof. Heymans' [...]

I am writing this letter while Mrs. Moll is reading aloud to her husband from the newspaper and I try to concentrate nevertheless; but we have never learned to do that under such circumstances, -- perhaps as we maybe need to, now that we have to live with many sheep in one pen.

[The letter ends with a long list of domestic issues that are not relevant here.]

Bye dear
\endmarg
\bigskip

Kapteyn indeed seems to have great difficulty sitting still. The work on this painting was done fairly concentrated in time at the location of the Astronomical Laboratory. The next letter a few days later has little on Kapteyn’s painting, but reports a narrow escape by Veth from serious injury or even death by  an accident in the street (my translation):
\begmarg
\hfill{Groningen Nov 27, 1917}

Misfortune is in a small corner and so is happiness. And so I ended up Sunday evening in great danger and miraculously escaped. I had dinner at the Kapteyns and going home I would take letters to the mailbox and therefore I made a slight detour because of a mailbox  I knew. But it was storming terribly and it was pitch black. Just when, having reached the box, wanting to put the letters in it, I heard a thunderous noise above me. And immediately afterwards I felt a terrible blow on my head and I fell to the ground feeling like a dead man. How I got up I don't know even now days later, but I had only one thought: to get away from that place, although I had the greatest difficulty walking and was trembling all over. I heard steps and was called after by someone who had my hat in his hand. He joined me and I held on to him and so I dragged myself to the door of the Moll family, while I was very distraught, but actually arrived unscathed. My support told me that I had received a top of a chimney on my head. I slept quite well and was only rather stiff the next morning. But after walking past the house where the accident had happened, I understood the danger I had escaped There are four large chimneys on the house, each equipped with a four-inch thick cover plate of more than a meter square in size and lined with lead. One of those four cover plates had been lifted up and had punched a large hole at the bottom of the sloping roof. There apparently it had tipped over and fell down along the gutter right on my head. Had I not, just before I went here, bought a new, \underline{sturdy} hat, which would have absorbed some of the impact (of course it was dented) then I probably would never have recounted all this. The blow would have landed right on top of my skull. Now it only drove me to the ground and I actually got off with only a fright, because at this moment, Tuesday evening, I feel very little of the stiffness that hindered me when painting yesterday. A girl who lives just across the mailbox told Miss Naardenburg, who lives in here [the maid?], that she was so startled by a terrible blow in the street, Sunday evening at ten o’clock, and that she could hardly believe that someone would survive  that blow. Also to me it is almost incomprehensible and yet I am in fact again quite all right again.

The painting becomes more and more fascinating the longer I work on it. I haven't produced anything so much mature in a long time. And now I want to make it really good. That is why I, while my intention was to stay here
no longer than fourteen days, I will work on it during the rest of this week.

And then I would try to be home by Saturday evening. But I'm not even sure, for a painting that has reached this stage everything else must give way. In any case, I think I stay at home then rather faithfully for the some time.

[The rest of the letter has domestic matters]

Bye dear Mom
\endmarg

So Veth considered the Kapteyn painting very special. And indeed it is. He felt an urge  to finish it as soon as he could and he was extending his stay in Groningen for that. We may conclude it had been  finished by November 1917 and then was  was presented to Mrs. Kapteyn three months later. What happened to the first portrait has not been mentioned anywhere in these letters. Most likely Veth kept it for himself.

\section{Veth's letters in 1921}

In 1921 Veth returned to Groningen. Although he did work on a painting of Kapteyn for the Senate Chamber, his reason for coming to Groningen was more than just that. He had been commissioned to produce a painting of another Groningen professor, Barend Sij\-mons (1853--1935), Professor of German Language and Literature. Sijmons and Kapteyn were very good friends. Indeed Sijmons features among the signatories of the Album accompanying the presentation of Kapteyn’s painting in 1918. There is an important parallel between the careers of both men (for a biography of  Sijmons and more background see de Wilde, 2006, 2007). The 1876 law on higher education not only gave rise to the appointment of Kapteyn as professor in astronomy, but also of Sijmons. Modern languages English, French and German were not taught at universities as an academic discipline at that time, nor as subjects at gymnasia. However they were part of the the curriculum of the HBS, because a career in  industry and commerce required a thorough knowledge of these languages. To provide for teachers necessary when the HBS was instituted, this higher education law stipulated that at least at one university education in these languages should be offered. However, not as full fledged academic studies, but only leading up to diploma’s for teaching qualification. This was remedied only in 1921(!), to a large extent due to efforts by Sijmons. The Groningen municipality saw this as an excellent manner of increasing the worryingly low student numbers. Three privaat-docents were to be appointed, with a salary provided by the municipality of Groningen. It started with Sijmons in 1878 for German (and until one was appointed also for English). This was a unique situation, since a privaat-docent was not supposed to be paid for his teaching, but merely appointed to lecture (usually a subject that was not yet taught).  Normally they had some other primary paid appointment at the university or elsewhere as well. In 1878 Sijmons taught at an HBS in the Frisian city of Sneek, but resigned when he was appointed and paid to teach in Groningen. This situation lasted not for long; in 1881 Sijmons was appointed full professor. So, Kapteyn and Sijmons had worked in Groningen over the same years. In 1921, two years before his retirement, Sijmons celebrated his fortieth anniversary as professor, and this painting was a made for that occasion, much like Kapteyn three years earlier. Unlike in Kapteyn’s case, however, the portrait offered at that celebration was meant for the Senate Chamber, so depicted Sijmons wearing academic gown, jabot with (in this case) bow-tie and beret (see Fig.~\ref{fig:Sijmons}).

\begin{figure}[t]
\sidecaption[t]
%\begin{center}
\includegraphics[width=0.60\textwidth]{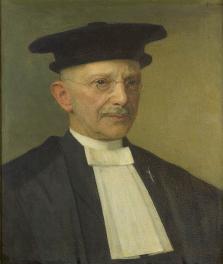}
%\end{center}
\caption{\normalsize Portrait of Barend Sij\-mons (1853--1935), professor of German language and literature, by Jan Veth. He would have been due to retire in 1923 at age seventy, but before that, in 1921, this painting was produced by Jan Veth on the occasion of the fortieth anniversary of his professorship. Oil on vanvas, $68\times 53.5$ cm. Courtesy University Museum Groningen.}
\label{fig:Sijmons}
\end{figure}

The Sijmons painting indeed is now displayed in the Senate Chamber. There were at least  two visits by Veth to Groningen. From the first one there are two letters to his wife (my translation):

\begmarg
 Groningen Thursday, Feb. 10, 1921

	It was a curious arrival here Tuesday evening. In fact the arrival at first was quite normal. We arrived exactly on time, and there was a carriage. Marie Vos was still up, - I got cup of tea, - and I went to bed fitfully. But at one o'clock I woke up with a dreadful pain. I thought it was a stomach ache, as I sometimes have. But I did not sleep for a moment that whole night and could have groaned if I had dared to. At eight o'clock I was finally called and the doctor was telephoned, who appeared at half past nine and ordered me to stay in bed. He first examined me and then gave me a strong narcotic powder and then an injection of morphine. But I could not go to sleep, although the pain slowly subsided. In the evening at eleven o'clock I fell asleep and slept very well that night. This morning I saw the doctor again, who told me I could get up, but taught me all kinds of cautionary lessons, and above all told me that I had to slow down and relax. I will have to seek advice from Dr. Beyerman.  The doctor here, who otherwise made a particularly serious impression, claims that it is gall stones, and that, if I live calmly, it is not a problem, but that if getting worse it may come necessary to undergo surgery. I do not intend to take the matter tragically right away. This morning I painted and this afternoon, with Isaac, I went to see the Israel Room in the museum. And I feel a little weak, but otherwise quite normal. A pleasant advantage is, that now, on medical authority, I can call  off things, that would make life too busy, which is welcome to me, […]. Soon I will write more.
\endmarg

I suppose Dr. Beyerman is Veth’s general practitioner. Marie Vos, with whom he is staying, is not identified. Veth wrote again two days later.
\begmarg
Groningen Feb. 12, 1921

	It is now Saturday evening, and I have had no more attacks, although I still feel some strange feelings of a more or less threatening nature in the lower abdomen all the time. But I am taking it easy, eating little and slowly, and devoting myself to the painting, that I am here for. I plan to return from here on Monday with the 4.36 train, in which there is a dining car, and can arrive in Bussum at 8.47. Then I will go to bed early,  […] Marie Vos, who has taken exceptional care of me, sends her regards.

Your Jan
\endmarg

His hostess Marie Vos is apparently known to Mrs. Veth as well. Note that he stated he is in Groningen for {\it the} painting, so singular. `Painting' cannot refer to the act of painting because in Dutch a different word would have been used. One would think this concerns Sijmons, since had it referred to Kapteyn he would probably have identified him. Isaac must be Isaac Lazarus Isra\"els (1865--1934), a Dutch painter belonging to the impressionists. In  the Veth archives in Dordrecht there are various letters between the two. The Museum for Ancient Art, the Predecessor of the current Groninger Museum, used to have an Isra\"els Room, in remembrance of Isaac's father  Jozef. Isaac Isra\"els and Veth had been planning to visit this room for some time according to their correspondence.

Veth ‘s health problems got worse and eventually he had to undergo surgery to remove the gallbladder and gallstones. This did not produce the results intended and he died in 1925 at age 61.
\bigskip

Later in February of 1921 Veth again visited Groningen, and there is one letter to his wife from this period (my translation):
\begmarg 
Bussum 25 Febr. 1921 (read: Groningen)

This evening a courier arrived with my proper address on the letter he brought, but instead of the letter probably meant for me the envelope had a letter inside to Joost. Although of course I immediately saw that it was addressed to Joost and not for me. I read it, because I still thought, that maybe you had  sent it on purpose to let me read it first. However, when I finished reading it, I realized that it was simply a mistake after all of putting the wrong letter in the wrong envelope. I will send it back to you, but would really want you not to send it on  in this form. For I have gradually come to the conclusion, carefully thinking about it, that the doctors must be mistaken, and that I did not have a gallstone attack, but actually really some kind of acute poisoning, which would leave nothing serious behind. Under these circumstances I would not like Joost to worry about me. Does a later outcome prove that the doctor was right, then we can always tell Joost.

I am working hard and doing very well. The portrait of Prof. Sijmons will probably not be finished, because after Sunday he probably won't be able to sit. But then I may still work on Monday and Tuesday on Prof. Kapteyn and therefore, since I am here anyway, I will stay until at least Tuesday evening. I will report later when I will get home.

With much love to Polle.

Your Jan
\endmarg

This is the only mention of Kapteyn in this period. Maybe the job was not a very time-consuming one. It does show that about six months before Kapteyn retired (at the end of the academic year 1920-21) Veth was back in Groningen working on a painting of Kapteyn, undoubtedly for the Senate Chamber, that was later offered to the University. Even if it were an overpainting of the first portrait with gown, jabot and beret, the careful Veth must have insisted that Kapteyn pose for it. There is nothing in this that contradicts Blaauw’s hypothesis.

\section{What can we make of all this?}

On the basis of these letters we know for certain that Jan Veth worked in 1917 on a first painting of Kapteyn in Groningen during a number of visits between May and July. He also produced a painting of Jan Willem Moll during these visits.  It is very likely that this first painting of Kapteyn that Veth produced in this period looked very closely similar to the preliminary version that is painted  on wood.

Immediately after this first painting had been finished, Veth started on a second painting with Kapteyn sitting at his desk. It seems unlikely that it was solely his decision, but it might have been in response to the opinion of Mrs. Kapteyn who disliked the first result of a stationary pose, preferring instead one in which he was shown  at work. Veth appeared delighted by the prospect of  producing a second painting of Kapteyn as he saw him as he approached his astronomical laboratory.

Who financed the new exercise is not clear; maybe Veth did not charge anyone for it. Kapteyn is wearing the same clothing at his desk as in the preliminary version and very likely also on the first painting, so Kapteyn simply continued posing in the same apparel. In November 1917 Veth returned to Groningen, during which time it was finished. It then was presented to Mrs. Kapteyn in February 1918 at the celebration and the jubilee of forty years professorship. Veth would have kept the first portrait and the preliminary version.

In 1921 Veth was in Groningen to produce a painting of Barend Sijmons and he did mention work on a paining of Kapteyn. This must have resulted in the Kapteyn painting in the Senate Chamber. The similarity in facial expression in the preliminary study and that in this portrait is striking; this supports the hypothesis of an overpainting of the first portrait with gown, jabot and beret. In addition,  the fact that the first painting no longer seems to exist is also easily explained this way. It would be likely anyway that when the second painting at the desk was the one going to be offered at Kapteyn’s professorship anniversary, Veth will have decided to keep the first one himself. And finally there is in addition to the facial expression the similarity in pose, expression and angle of view between the preliminary version and the one in the Senate Chamber, even in details like the position of eyelids, direction in which Kapteyn is looking, etc. Had Veth produced in 1921 an independent portrait starting again from scratch with Kapteyn posing again for many hours there would unavoidably have been more than minor differences.

The only piece of evidence that points away from Blaauw’s hypothesis is that the director of the University Museum, Lars Hendrikman, while I was present, carefully examined the painting from the Senate Chamber and concluded that it shows no sign of an overpainting of the gown, jabot and beret. UV lightning did not show color changes due to the background having been varnished twice and the gown, etc. only once. Definite confirmation on this should be obtainable by  an examination with modern imaging techniques such using  X-ray or infrared radiation, or other electromagnetic radiation or bundles of  elementary particles. Unfortunately this is not feasible without dedicated funding. Blaauw’s remark that traces of hair betrayed Veth’s overpainting exercise were not corroborated. But then it seems to me against Veth’s extreme care for details and perfectionist approach to leave such traces in the first place. That he signed the painting with the year 1921 is also understandable, even if it was produced by overpainting, since after all it had been finished that year.

A very strong piece of evidence, corroborating the Blaauw overpainting scenario,  is the listing of Veth’s paintings in the biography by Veth’s good friend  Johan Huizinga. The painting now in the Senate Chamber is dated {\it before} the painting of Kapteyn behind his desk. Huizinga should be well informed. The dating fits the notion that the Senate portrait is produced by overpainting  the original from 1917 with academic attire. Huizinga lists only two  paintings; if the Blaauw hypothesis is false there should be a third  painting in the Huizinga listing, produced in 1921 and meant for the Senate chamber. Huizinga's dating and absence of a third painting constitute quite strong evidence; he himself described his intimate familiarity with the making of some paintings of professors, including the ones of Kapteyn.
\bigskip

The most consistent scenario seems to me the following.
\begin{itemize}
\item In the spring of 1917 Jan P. Veth was asked to paint two portraits in Groningen: one  of Jan Willem Moll to be offered to him after his resignation as ordinary professor, and one of Jacobus C. Kapteyn on the occasion of his jubilee of 40 years as professor in February 1918. 
\item Veth picked this up by painting in Groningen in the home of the Moll family, where both Moll and Kapteyn sat. 
\item Of Kapteyn he made a first design on a wooden panel, which he kept for himself. He gave it later to his friend and colleague Georg Rueter and is now in the possession of Kapteyn’s great-grandson Jack (also Jacobus Cornelius).  
\item The painting by Moll shows the latter sitting with his arms crossed in ordinary clothing, so was also not intended for the Senate Chamber. 
\item The first painting of Kapteyn must have looked very much like the preliminary study, with Kapteyn wearing ordinary clothes, since it likewise was not intended for the Senate Chamber.
\item After the paintings were finished, Veth immediately began a second one showing Kapteyn sitting behind his desk. So it was probably produced without a fee, quite possibly because Mrs. Kapteyn, to whom the portrait was to be presented, was not satisfied with the first portrait, as Adriaan Blaauw learned from Pieter van Rhijn.
\item Veth was delighted with the prospect, because it offered a far more attractive perspective than the ‘standard’ portrait. It shows Kapteyn in the same clothes as in the preliminary study, who would have agreed to sit for the painting since it was to please his wife.
\item It was finished in November 1917 and in February 1918 the second painting was offered at the jubilee celebration, according to the Rector's remarks who hoped it would be offered later to the university in spite of the differed setting and format compared to the professors gallery in the Senate Chamber. This scenario fully agrees with Johan Huizinga's listing of Veth's works. 
\item In February 1921 Veth was back in Groningen to make a portrait for the Senate Chamber of Barend Sijmons as well as one of Kapteyn for the same destination. 
\item Of Kapteyn's first portrait nothing was ever seen again. Veth paintings do not get lost that easily one would think. A very straightforward explanation would be that during this  1921 episode in Groningen it was overpainted with gown, jabot and beret. Kapteyn must have been pleased that he then did not have to sit and pose for it for very long hours. This explains Huizinga's statement that the first painting produced in 1917 ended up in the Senate Chamber.  
\item We can infer from the preliminary study what the original portrait would have looked like before the repainting. The striking similarity with the painting in the Senate Chamber in pose, expression, angle of view and many details support this course of events.
\end{itemize}
}

%&*&

\noindent
\section{Acknowledgments}

{\normalsize
I am grateful in the first place to the late Adriaan Blaauw for bringing this up at the 1999 Legacy Symposium. Throughout the years Klaas van Berkel, Professor of History and co-organizer of that symposium, and Franck Smit, former Curator and Director of the Groningen University Museum, have encouraged my interest and supported my research on historical matters with advice and suggestions. This paper has profited from remarks, suggestions and help, in addition of these two individuals, by Arjan Dijkstra, former Director, Sieger Vreeling, Curator collections, and Lars Hendrikman, current Director of the Groningen University Museum. I am also grateful to Quirine van der Meer Mohr, curator 19th century of the Dordrechts Museum, for some detailed correspondence and suggestions.  I am very grateful to Klaas van Berkel and Franck Smit for critically reading a draft of this paper. I thank the staff of the Kapteyn Astronomical Institute and the Director, Professor L\'eon Koopmans, for support and help, particularly for asking and financing me and my wife to represent the Institute at the opening of the Veth exhibition in Dordrecht, and in general for hospitality extended to an Emeritus Professor as Guest Scientist.
\vspace{0.7cm}
}

%&*&
\section{References}
{\normalsize

\ \    Bijl de Vroe, F.,  1987. {\it De schilder Jan Veth, 1864--1925: Chroniqueur van een bewogen tijdperk}. Am\-ster\-dam: Thomas Rap. ISBN  978-90-60052-532.

Bijl de Vroe, F., van den Donk, C., Ekkart, R.,  van der Meer Mohr, Q., Reid, N. \&\ Rens., A., 2023. {\it Het oog van Jan Veth: Schilder en criticus rond 1900}. Zwolle: Waanders. ISBN 978-94-62624-313.  

de Sitter, W., 1925. A biography of Kapteyn, {\it The Observatory}, 48, 293-294.

de Wilde, I., 2006. De pionier van het moderne talenonderwijs: Leven en werk van Barend Sijmons. {\it Biografie Bulletin}, najaar 2006, 55-58. Online: www.dbnl.org/tekst/\_bio001200601\_01/\_bio001200601\_01\_\linebreak[4]
0050.php.

de Wilde, I., 2007. {\it Werk maakt het bestaan draaglijk: Barend Sijmons (1853-1935)}, Groningen, Barkhuis Publishing, ISBN 978-90-77922-293.

Einstein, A. \&\ de Sitter, W., 1932. On the relation between  the expansion and the  mean density of the Universe, {\it Proceedings of the National Academy of Sciences of the United States of America}, 18, 213–214.

Gill, D. \&\ Kapteyn, J.C. 1896, 1897, 1900. The Cape Photographic Durchmusterung for the equinox 1875. Part I. Zones -18\degs\  to -37\degs; Part II. Zones -38\degs\  to -52\degs; Part III. Zones -53\degs\  to -89\degs. {\it Annals of the Cape Observatory, South Africa}, 3, 1–845; 4, 1–702;  5, 1–757.

Hertzsprung-Kapteyn, H., 1928. {\it J.C. Kapteyn; Zijn leven en werken}. Groningen, Wolters (electronic version, available at www.dbnl.org/tekst/hert042jcka01\_01/; for more information and my English translation see www.astro.rug.nl/JCKapteyn/HHKbiog.html).

Huizinga, J., 1927. {\it Leven en werk van Jan Veth}, Haarlem: Tjeenk Willink \&\ Zoon. 

Kapteyn, J.C. 1906. {\it Plan of Selected Areas}. Astronomical Laboratory at Groningen, Hoitsema Brothers, Groningen.

Kapteyn, J.C. 1914. Het sterrekundig laboratorium. In: {\it Academia Groningana MDCXIV-MCMXIV: Gedenkboek ter gelegenheid van het derde eeuwfeest der Universiteit te Groningen}. University of Groningen, 550–552.

Kapteyn, J.C. \&\ van Rhijn, P.J., 1920. On the distribution of the stars in space especially in the high Galactic latitudes. {\it Astrophysical Journal}, 52, 23–38.

Kapteyn, J.C., 1922. First attempt at a theory of the arrangement and motion of the Sidereal System. {\it Astrophysical Journal}, 55, 302–328.

Lucas, F.L.R. \&\ van der Salm, S.A.M., 2017.  Replicating the lost skew distribution machine of Jacobus Cornelius Kapteyn, In: {\it Calculating in everyday life: Slide rules, tables, calculators and other mathematical devices}, ed. K. Kleine, Deutsche Nationalbibliothek, ISBN 978-3-7448-1056-2, 231-267. 

Oort, J. H. 1927. Observational evidence confirming Lindblad’s hypothesis of a rotation of the Galactic System. {\it Bulletin of the Astronomical Institutes of the Netherlands}, 3, 275–282. 

Oort, J.H., 1938. Absorption and density distribution in the Galactic System. {\it Bulletin of the Astronomical Institutes of the Netherlands}, 8, 233–264.

Oosterheert, J. (ed.), 2009. {\it In vol ornaat: Vier eeuwen Senaatsgalerij}. Groningen; Universiteitsmuseum.

Regionaal Archief Dordrecht, 2022. See www.regionaalarchiefdordrecht.nl/archive/search/mivast=46\linebreak[4]
mizig=301\&miadt=301\&milang=en\&mizk\_all=jan\%20veth\&miview=ldt.

University of Groningen, 1918. {\it Jaarboek der Rijksuniversiteit Groningen 1917-1918}. Groningen, J.B. Wolters. 

University of Groningen, 1921. {\it Jaarboek der Rijksuniversiteit Groningen 1920-1921}. Groningen, J.B. Wolters. 

University of Groningen, 2022. {\it Mural of J.C. Kapteyn} (Website), see www.rug.nl/museum/exhibitions/\linebreak[4]
permanent/mural-of-j-c-kapteyn.

van der Kruit, P.C. \&\ van Berkel, K., 2000. {\it The legacy of J.C. Kapteyn: Studies on Kapteyn and the development of modern astronomy}. Dordrecht, Kluwer, ISBN 978-94-010-9864-9.

van der Kruit, P.C., 2015. {\it Jacobus Cornelius Kapteyn: Born investigator of the Heavens}. Springer, ISBN 978-3-319-10875-9, accompanying Webpage www.astro.rug.nl/JCKapteyn.

van der Kruit, P.C., 2019. {\it Jan Hendrik Oort: Master of the Galactic System}, Springer, ISBN 978-3- 319-10875-9, accompanying Webpage www.astro.rug.nl/JHOort.

van der Kruit, P.C., 2021a. {\it Pioneer of Galactic Astronomy: A biography of Jacobus Cornelius Kapteyn}. Springer Publishers. ISBN 978-3-030-55422-4.

van der Kruit, P.C., 2021b. {\it Master of Galactic Astronomy: A biography of Jan Hendrik Oort}, Springer, ISBN 978-3-030-55547-4.

van der Kruit, P.C., 2021c. Karl Schwarzschild, Annie J. Cannon and Cornelis Easton: PhDs Honoris Causa of Jacobus C. Kapteyn. {\it Journal for Astronomical History and Heritage}, 24, 521–543.

van der Kruit, P.C., 2022a. Pieter Johannes van Rhijn, Kapteyn’s Astronomical Laboratory and the Plan of Selected Areas. {\it Journal of Astronomical History and Heritage}, 25, 341-438.

van der Kruit, P.C., 2022b. {\it J.C. Kapteyn en zijn Sterrenkundig Laboratorium een eeuw later -- J.C. Kapteyn and his Astronomical Laboratory a centurty later}, Kapteyn Astronomical Institute, ISBN 978-90-903-6008-9. Available only electronically at www.astro.rug.nl/~vdkruit/JCK100.pdf.

van der Kruit, P.C., 2023. Excited states and spontaneous transitions: Astronomer, lecturer, administrator, biographer. {\it Journal of Astronomical History and Heritage}, 26, 203-25.

Willink, B., 1991. Origins of the second Golden Age of Dutch science after 1860, {\it Social studies of science}, 21, 389–413.
}

\end{document}